\documentclass[%
 reprint,
nofootinbib,
 amsmath,amssymb,
onecolumn,
11pt,
floatfix,
showkeys
]{revtex4-2}

\usepackage{graphicx}
\usepackage{dcolumn}
\usepackage{bm}
\usepackage[colorlinks]{hyperref}
\usepackage{wasysym}

\newcommand{\enterprise}{\texttt{ENTERPRISE}}
\newcommand{\GWecc}{\texttt{GWecc}}
\newcommand{\GWeccjl}{\texttt{GWecc.jl}}

\allowdisplaybreaks

\begin{document}

\preprint{APS/123-QED}

\title[]{Post-Newtonian-accurate pulsar timing array signals induced by inspiralling eccentric binaries: accuracy, computational cost, and single-pulsar search}

\author{Abhimanyu Susobhanan}
\email{abhimanyu.susobhanan@nanograv.org} 
 \affiliation{Center for Gravitation, Cosmology, and Astrophysics, University of Wisconsin-Milwaukee, Milwaukee, WI 53211, USA} 

 




\date{\today}

\begin{abstract}
Pulsar Timing Array (PTA) experiments are expected to be sensitive to gravitational waves (GWs) emitted by individual supermassive black hole binaries (SMBHBs) inspiralling along eccentric orbits.
We compare the computational cost of different methods of computing the PTA signals induced by relativistic eccentric SMBHBs, namely approximate analytic expressions, Fourier series expansion, post-circular expansion, and numerical integration.
We show that the fastest method for evaluating PTA signals is by using the approximate analytic expressions, {which provides} up to $\sim$50 times improvement in computational {speed} over the alternative methods.
We investigate the accuracy of the approximate analytic expressions by employing a mismatch metric valid for PTA signals. 
We show that this method is accurate within the region of the binary parameter space that is of interest to PTA experiments.
We introduce a spline-based {method to further accelerate} the PTA signal evaluations for narrowband PTA datasets.
{The efficient methods for computing the eccentric SMBHB-induced PTA signals were implemented in the \GWeccjl{} package and can be readily accessed from the popular \enterprise{} package to search for such signals in PTA datasets.
Further, we simplify the eccentric SMBHB PTA signal expression for the case of a single-pulsar search and  demonstrate our computationally efficient methods by performing a single-pulsar search in the 12.5-year NANOGrav narrowband dataset of PSR J1909$-$3744 using the simplified expression.
These results will be crucial for searching for eccentric SMBHBs
in large PTA datasets.
}
\end{abstract}


\maketitle


\section{Introduction}

Routine detections of gravitational waves (GWs) from stellar mass compact binary merger events by ground-based GW detectors {such as the advanced LIGO and the advanced Virgo} have ushered in the era of GW astronomy \cite{AbbotAbbott+2019,AbbotAbbott+2021a,AbbotAbbot+2021b}. 
Pulsar Timing Arrays \cite[PTAs:][]{Sazhin1978,FosterBacker1990} are experiments that aim to detect GWs in the nanohertz frequency range by routinely monitoring ensembles of millisecond pulsars using some of the world's most sensitive radio telescopes.
Ongoing PTA campaigns include the North American Nanohertz Observatory for Gravitational waves \cite[NANOGrav:][]{AlamArzoumanian+2020}, the European Pulsar Timing Array \cite[EPTA:][]{DesvignesCaballero+2016}, the Parkes Pulsar Timing Array \cite[PPTA:][]{KerrReardon+2020}, the Indian Pulsar Timing Array \cite[InPTA:][]{TarafdarNobleson+2022}, {the MeerKAT Pulsar Timing Array} \cite[MPTA:][]{MilesShannon+2022}, and the Chinese Pulsar Timing Array \cite[CPTA:][]{Lee2016}.
The International Pulsar Timing Array consortium \cite[IPTA:][]{PereraDeCesar+2019} aims to combine data and resources from different PTA campaigns to accelerate the discovery of nanohertz GWs and improve the prospects of post-discovery science.
PTA experiments have grown in sensitivity over the years, and are expected to achieve their first detection in the near future \cite{GoncharovShannon+2021,ChenCaballero+2021,ArzoumanianBaker+2020b,AntoniadisArzoumanian+2022}.

The most prominent sources of nanohertz GWs are expected to be supermassive black hole binaries (SMBHBs), usually hosted by active galactic nuclei (AGNs) \cite{Burke-SpolaorTaylor+2019}.
The first PTA detection is expected to be that of a stochastic GW background \citep[GWB:][]{HellingsDowns1983} formed via the incoherent addition of GWs emitted by an ensemble of unresolved SMBHBs, followed by the detection of individual SMBHBs that stand out above the GWB \cite{KelleyBlecha+2017,KelleyBlecha+2018,PolTaylor+2021}.
Several promising SMBHB candidates have been identified through electromagnetic observations of AGNs \cite[e.g.][]{IguchiOkudaSudou2010,DeyGopakumar+2019,HuDOrazio+2020,XinMingarelliHazboun2021}, and PTA experiments have already put increasingly stringent constraints on the presence of SMBHB signals in their datasets \cite{JenetLommen+2004,BabakPetiteau+2016,FengLi+2019,AggarwalArzoumanian+2019,ArzoumanianBaker+2020}.

SMBHBs are believed to form through galaxy mergers, where the central black holes of the merging galaxies sink to the center of the merger remnant, eventually forming a bound system \cite{BegelmanBlandfordRees1980}.
Such binary systems shrink due to energy and angular momentum exchange with the surrounding stars and gas until the orbital evolution is dominated by GW emission \cite{DottiSesanaDecarli2012}, and can retain significant eccentricities as they enter the PTA frequency band  \cite[e.g.][]{ArmitageNatarajan2005,RoedigSesana2012}.
The SMBHB candidate OJ 287 is believed to host a binary system with eccentricity $\sim0.6$ \cite{DeyGopakumar+2019}.
It is therefore desirable to search for signals induced by eccentric SMBHB systems in PTA datasets.

The PTA responses to GW signals (known as PTA signals) due to inspiralling eccentric SMBHB systems were modeled by Refs.~\cite{JenetLommen+2004,TaylorHuerta+2016,SusobhananGopakumar+2020} (see Section \ref{sec:pta-signal} for details).
This usually involves modeling the relativistic motion of the binary system using the post-Newtonian (PN) formalism, where general relativistic corrections to Newtonian dynamics are expressed in powers of $(v/c)^2\sim GM/(c^2 r)$, and where $M$ is the total mass of the binary, $r$ is the relative separation, and $v$ is the relative speed \cite{Blanchet2014}.
The GW strain amplitudes in the two orthogonal polarizations $h_{+,\times}$ can then be expressed in terms of the polar coordinates $r$ and  $\phi$ in the orbital plane and their time derivatives.
Finally, the PTA signal $R(t)$ involves the time integrals of $h_{+,\times}(t)$.

A significant challenge in searching for such signals in PTA datasets is posed by the cost of computing the PTA signal given a set of pulse times of arrival (TOAs) of an ensemble of pulsars.
(See Refs~\cite{TaylorHuerta+2016,ZhuWen+2015} for estimates of the sensitivity degradation experienced when using the computationally inexpensive circular PTA signals \cite{JenetLommen+2004} to search for eccentric sources.)
This computational cost is mainly incurred in two stages: (a) solving the orbital evolution, and (b) computing the PTA signal as a function of the orbital variables.
Refs~\cite{MooreRobson+2018,SusobhananGopakumar+2020} provided an analytic solution to the quadrupolar-order orbital evolution equations governing the motion of inspiralling non-spinning eccentric binaries, thereby mitigating the computational cost of solving the orbital evolution of such systems.
A few different approaches have been presented in the literature for computing the PTA signals given the orbital variables as a function of time, including an approximate analytic integral  \cite{JenetLommen+2004}, a Fourier series expansion \cite{TaylorHuerta+2016}, a post-circular expansion, and numerical integration \cite{SusobhananGopakumar+2020}.
In this work, we compare the computational cost of these methods and investigate the accuracy of the most efficient method.
We then introduce a new spline-based method for further improving the computational performance of PTA signal evaluations by leveraging the properties of real PTA datasets.
{We have implemented the optimal methods for computing the PTA signals due to eccentric SMBHBs in the \GWeccjl{} package, and this can be used readily with the popular \enterprise{}  package \cite{EllisVallisneri+2020} for searching PTA datasets.
Further, to demonstrate our methods, we perform a single-pulsar search for eccentric SMBHB signals in the NANOGrav 12.5-year narrowband dataset of PSR J1909$-$3744 \cite{AlamArzoumanian+2020}.}

This paper is arranged as follows. 
In Section \ref{sec:pta-signal}, we introduce the PTA signals induced by inspiralling eccentric binaries, and the GW phasing approach used for their computation.
We compare the computational cost of different approaches for evaluating the PTA signals given the orbital evolution as a function of time in \ref{sec:comp-cost}.
We investigate the accuracy of the most computationally efficient PTA signal computation method in \ref{sec:spx_anl_accuracy}.
In Section \ref{sec:spline-method}, we introduce a new spline-based {method to further reduce} the computational cost of evaluating PTA signals by exploiting the structure of typical PTA datasets.
{We provide a brief description of the \GWeccjl{} package in Section \ref{sec:gwecc.jl}.
In Section \ref{sec:1psr-search}, we demonstrate our methods by performing a single-pulsar search for eccentric SMBHB signals using the NANOGrav 12.5-year narrowband data of PSR J1909$-$3744 as a proof of concept.}
Finally, we summarize our results in Section \ref{sec:summary}. 

\section{The PTA signal model for eccentric binaries}
\label{sec:pta-signal}

\subsection{The gravitational waveform and the PTA signal}

We begin by briefly describing the PN-accurate PTA signal model for inspiralling eccentric binaries.
GWs traveling across the line of sight to a pulsar {induce} modulations on the observed TOAs of its pulses.
These modulations are given by {\cite{EstabrookWahlquist1975,Detweiler1979,JenetLommen+2004}}
\begin{equation}
R(t_E)=\int_{t_{0}}^{t_E}\left(h(t_E')-h(t_E'-\Delta_{p})\right)dt_E'\,,
\label{eq:pta-res}
\end{equation}
where $h$ is the GW strain (precisely defined below), the time variables $t_E$ and $t_E'$ are measured in the solar system barycenter (SSB) frame, $\Delta_{p}$ is a geometric time delay given by
\begin{equation}
\Delta_{p}=\frac{D_{p}}{c}\left(1-\cos\mu\right)\,,
\end{equation}
$D_{p}$ is the pulsar distance, $\mu$ is the angle between the lines of sight to the pulsar and the GW source, {$c$ is the speed of light}, and $t_0$ is an arbitrary fiducial time. 
The coordinate time $t_E$ measured in the SSB frame relates to the coordinate time $t$ measured in the GW source frame via the cosmological redshift as
\begin{equation}
    t_E - t_0 = (1+z)(t-t_0)\,.
    \label{eq:redshift}
\end{equation}
{Despite the above distinction between the coordinate times measured in the SSB frame ($t_E$) and the GW source frame ($t$), the redshift $z$ is not measurable for an individual SMBHB source due to the fact that it is covariant with the orbital frequency and the total mass of the source \cite[see, e.g.][]{MessengerTakami+2014}.
Therefore, we drop the subscript $E$ denoting the SSB frame with the understanding that the frequencies and masses appearing in the equations hereafter have been scaled by redshift.}


The dimensionless GW strain $h(t)$ is given by
\begin{equation}
h(t)=\begin{bmatrix}F_{+} & F_{\times}\end{bmatrix}\begin{bmatrix}\cos2\psi & -\sin2\psi\\
\sin2\psi & \cos2\psi
\end{bmatrix}\begin{bmatrix}h_{+}(t)\\
h_{\times}(t)
\end{bmatrix}\,,
\end{equation}
where $h_{+,\times}(t)$ are the two GW polarization amplitudes, $F_{+,\times}$ are the antenna pattern functions that depend on the sky locations of the pulsar and the GW source (see, e.g., Ref.~\cite{LeeWex+2011} for explicit expressions), and $\psi$ is the GW polarization angle. 
Hence, $R(t)$ can be expressed in terms of functions  $s_{+,\times}(t)$ defined as
\begin{equation}
s_{+,\times}(t)=\int_{t_{0}}^{t}h_{+,\times}(t')dt'\,,
\end{equation}
such that 
\begin{equation}
R(t)=\begin{bmatrix}F_{+} & F_{\times}\end{bmatrix}\begin{bmatrix}\cos2\psi & -\sin2\psi\\
\sin2\psi & \cos2\psi
\end{bmatrix}\begin{bmatrix}s_{+}(t)-s_{+}(t-\Delta_{p})\\
s_{\times}(t)-s_{\times}(t-\Delta_{p})
\end{bmatrix}\,.
\label{eq:R(t)_px}
\end{equation}
The $s_{+,\times}(t)$ and $s_{+,\times}(t-\Delta_p)$ contributions are known as the Earth term and the pulsar term respectively.

The quadrupolar-order $h_{+,\times}$ expressions valid for binary systems inspiralling along eccentric orbits are given by \cite{SusobhananGopakumar+2020}
\begin{subequations}
\begin{align}
h_{+}&=H\left[\left(c_{\iota}^{2}+1\right)\left(\frac{\left(2e_{t}^{2}-\chi^{2}+\chi-2\right)}{(1-\chi)^{2}}\cos(2\phi)-\frac{2\sqrt{1-e_{t}^{2}}\xi}{(1-\chi)^{2}}\sin(2\phi)\right)+s_{\iota}^{2}\frac{\chi}{(1-\chi)}\right]\,,\\
h_{\times}&=H\,2c_{\iota}\left[\frac{2\sqrt{1-e_{t}^{2}}\xi}{(1-\chi)^{2}}\cos(2\phi)+\frac{\left(2e_{t}^{2}-\chi^{2}+\chi-2\right)}{(1-\chi)^{2}}\sin(2\phi)\right]\,,
\end{align}
\label{eq:hpx}
\end{subequations}
where 
$H=\frac{GM\eta}{D_{L}\,c^{2}}x$, 
$\eta$ is the symmetric mass ratio, 
$D_{L}$ is the luminosity distance to the binary, 
$x=\left(GM(1+k)n/c^{3}\right)^{2/3}$ is a dimensionless PN parameter, 
$n$ is the mean motion of the orbit, 
$k$ is the relativistic advance of periapsis per orbit, 
$e_{t}$ is the time eccentricity, 
$c_{\iota}=\cos\iota$, 
$s_{\iota}=\sin\iota$,
$\iota$ is the orbital inclination, 
$\chi=e_{t}\cos u$, 
$\xi=e_{t}\sin u$, 
$u$ is the true anomaly, and 
$\phi$ is the angular coordinate in the orbital plane also known as the orbital phase. 
The variables $\phi$ and $u$ can be obtained as functions of time using the gravitational wave phasing approach \cite{DamourGopakumarIyer2004}, and this is what we discuss in the next subsection. 

\subsection{The GW phasing formalism for computing the orbital motion}
\label{sec:phasing}

The conservative dynamics of the binary system can be expressed using the PN-accurate quasi-Keplerian parametrization \cite{DamourDeruelle1985,MemmesheimerGopakumarSchafer2004}.
We begin by defining the mean anomaly $l$ as
\begin{equation}
l(t)=\int_{t_{0}}^{t}n(t')dt'\,.
\end{equation}
The eccentric anomaly $u$ can be written implicitly as a function of $l$ using the PN-accurate Kepler equation \cite{MemmesheimerGopakumarSchafer2004,BoetzelSusobhanan+2017}
\begin{equation}
l=u-e_{t}\sin u+\mathfrak{F_{t}}(u)\,,
\label{eq:kepler}
\end{equation}
where $e_t$ is known as the time eccentricity and $\mathfrak{F_{t}}(u)$ is a periodic function of $u$.
Although this transcendental equation admits an analytic Fourier series solution \cite{BoetzelSusobhanan+2017}, it is usually solved numerically in practice due to the computationally expensive nature of the analytic solution.
An efficient numerical method for solving the Newtonian Kepler equation was introduced by Ref.~\cite{Mikkola1987} and can be adapted to solve the PN-accurate Kepler equation. 

The orbital phase $\phi$ can be written as
\begin{equation}
\phi=\gamma+l+(1+k)(f-l)+\mathfrak{F}_{\phi}(u)\,,
\end{equation}
where $f$ is the true anomaly given by
\begin{equation}
f=2\arctan\left[\sqrt{\frac{1+e_{\phi}}{1-e_{\phi}}}\tan\frac{u}{2}\right]\,,
\end{equation}
$k$ is the advance of periapsis per orbit, 
$\gamma$ is the angle of periapsis defined as 
\begin{equation}
\gamma(t)=\int_{t_{0}}^{t}k(t')n(t')dt'\,,
\end{equation}
$e_{\phi}$ is known as the angular eccentricity, and $\mathfrak{F}_{\phi}(u)$  is a periodic function of $u$.
The periastron angle $\gamma$ is not to be confused with the argument of periastron $\omega$, which is defined as 
$\omega=\phi-f$; instead, $\gamma(t)$ represents the orbital-averaged secular evolution of $\omega$.
Explicit expressions for $k$, $e_{\phi}$, $\mathfrak{F_{t}}(u)$ and $\mathfrak{F}_{\phi}(u)$ in terms of $e_{t}$, $n$, $M$ and $\eta$ can be found in, e.g., Ref.~\cite{MemmesheimerGopakumarSchafer2004}. 
In the Newtonian limit $x\rightarrow0$, we have 
$e_{\phi},e_{t}\rightarrow e$, 
$k\rightarrow0$, 
$\mathfrak{F}_{t}\rightarrow0$,
$\mathfrak{F}_{\phi}\rightarrow0$, and
$\omega\rightarrow\gamma$. 

The reactive evolution of the orbit due to gravitational radiation reaction can be incorporated into this formalism by allowing $n$ and $e_{t}$ to vary slowly with time. 
Up to the leading quadrupolar order, this can be expressed as a system of four coupled ordinary differential equations (ODEs) \cite{DamourGopakumarIyer2004,SusobhananGopakumar+2020}
\begin{subequations}
\begin{align}
\frac{dn}{dt} & =\frac{1}{5}\left(\frac{GMn}{c^{3}}\right)^{\frac{5}{3}}\eta n^{2}\frac{\left(96+292e_{t}^{2}+37e_{t}^{4}\right)}{\left(1-e_{t}^{2}\right)^{7/2}}\,,\\
\frac{de_{t}}{dt} & =\frac{-1}{15}\left(\frac{GMn}{c^{3}}\right)^{\frac{5}{3}}\eta ne_{t}\frac{\left(304+121e_{t}^{2}\right)}{\left(1-e_{t}^{2}\right)^{5/2}}\,,\\
\frac{d\gamma}{dt} & =k\,n\,,\\
\frac{dl}{dt} & =n\,.
\end{align}
\label{eq:orbit-odes}
\end{subequations}
An analytic solution for this set of equations was derived by Refs.~\cite{MooreRobson+2018,SusobhananGopakumar+2020}, and involves hypergeometric functions.
(We do not provide this solution here explicitly due to its length.)
This solution allows us to evolve the binary orbit efficiently without the need for numerical ODE solvers.


\subsection{Computing the PTA signal}
\label{sec:PTA-signal-compute}

Given the complicated PN-accurate functional forms of $h_{+,\times}(t)$, it is in general not possible to obtain exact expressions for their integrals $s_{+,\times}(t)$, which are necessary for computing the PTA signal $R(t)$.
Four different approaches for computing $s_{+,\times}(t)$ given the dynamic variables $n(t)$, $e_t(t)$, $\gamma(t)$, $l(t)$ have been presented in the literature, and these methods are listed below.

~

\textbf{Method 1 (\textit{Analytic})}: 
Ref.~\cite{JenetLommen+2004} derived analytic expressions for $s_{+,\times}(t)$ assuming slow advance of periapsis and orbital decay compared to the orbital motion, as well as the Newtonian limit for the eccentricities $e_\phi\rightarrow e_t$.
These expressions, re-written in terms of $u$ and $\omega$ in our notation, are given by
\begin{subequations}   
\begin{align}
s_{+}^\text{A}(t)&=\frac{H}{n}\left(\left(c_{\iota}^{2}+1\right)\left(-\mathcal{P}\sin(2\omega)+\mathcal{Q}\cos(2\omega)\right)+s_{\iota}^{2}\mathcal{R}\right)\,,\\
s_{\times}^\text{A}(t)&=\frac{H}{n}2c_{\iota}\left(\mathcal{P}\cos(2\omega)+\mathcal{Q}\sin(2\omega)\right)\,,
\end{align}
\label{eq:spx_anl}
\end{subequations}  
where
\begin{subequations}
\begin{align}
\mathcal{P} &=\frac{\sqrt{1-e_t^{2}}(\cos(2u)-e_t\cos(u))}{1-e_t\cos(u)}\,,\\
\mathcal{Q} &=\frac{\left(\left(e_t^{2}-2\right)\cos(u)+e_t\right)\sin(u)}{1-e_t\cos(u)}\,,\\
\mathcal{R} &= e_t \sin(u)\,.
\end{align}
\label{eq:PQ_int}
\end{subequations}
Specifically, $s_{+,\times}^\text{A}(t)$ obey the relation
\[
h_{+,\times}={n}\;\frac{\partial s_{+,\times}^{\text{A}}}{\partial l}\Bigg|_{e_\phi\rightarrow e_t;n,e_{t},\omega}\,.
\]
Note that these expressions are exact when only the Newtonian motion is considered, where $\omega$ is a constant. 
In the case of relativistic binaries with a non-zero advance of periapsis, the PN-accurate values of $u$ and $\omega$ can be substituted.

~

\textbf{Method 2 (\textit{Fourier})}: 
Ref.~\cite{TaylorHuerta+2016} derived Fourier series expansions for $s_{+,\times}(t)$ using the Fourier series expansions of $h_{+,\times}(t)$ \cite{PetersMathews1963}, assuming slow advance of periapsis and orbital decay, as well as the Newtonian limit for the eccentricities $e_\phi\rightarrow e_t$, and neglecting the orbital-timescale variations of $\omega$.
In our notation, these expressions can be written as
\begin{subequations}
\begin{align}
   s_{+}^\text{F}(t)&=\frac{H}{n}\sum_{p=1}^{\infty}\Bigg[\left(c_{\iota}^{2}+1\right)\Big(-\overline{\mathcal{P}}\cos(pl)\sin(2\gamma)+\bar{\mathcal{Q}}\sin(pl)\cos(2\gamma)\Big)+s_{\iota}^{2}\bar{\mathcal{R}}\sin(pl)\Bigg]\,,\\
   s_{\times}^\text{F}(t)&=\frac{H}{n}2c_{\iota}\sum_{p=1}^{\infty}\Big(\overline{\mathcal{P}}\cos(pl)\cos(2\gamma)+\bar{\mathcal{Q}}\sin(pl)\sin(2\gamma)\Big)\,,
\end{align}
\label{eq:spx_pm}
\end{subequations}
where
\begin{subequations}
\begin{align}
\bar{\mathcal{P}}&=\sqrt{1-e_t^{2}}\left[J_{p-2}(pe_t)+J_{p+2}(pe_t)-2J_{p}(pe_t)\right]\,,\\
\bar{\mathcal{Q}}&=-J_{p-2}(pe_t)+2e_t J_{p-1}(pe_t)-\frac{2}{p}J_{p}(pe_t)-2e_t J_{p+1}(pe_t)+J_{p+2}(pe_t)\,,\\
\bar{\mathcal{R}}&=\frac{2}{p}J_{p}(pe_t)\,,
\end{align}
\end{subequations}
and $J_n(x)$ represent Bessel functions of the first kind.
It is straightforward to show that the above Fourier series expansions $s_{+,\times}^\text{F}(t)$ are equivalent to Eqs.~\eqref{eq:spx_anl} in the limit $\omega\rightarrow\gamma$.

~

\textbf{Method 3 (\textit{Post-circular})}:
Ref.~\cite{SusobhananGopakumar+2020} derived a post-circular expansion (expansion in powers of $e_t$) for $s_{+,\times}(t)$ valid for low eccentricities ($e_t<0.3$) incorporating fully 3PN-accurate advance of periapsis using the post-circular expansion of $h_{+,\times}(t)$ derived in Ref.\cite{BoetzelSusobhanan+2017}.
This expansion has the form
\begin{align}
  s_{+,\times}^{\text{PC}}(t)&=\frac{H}{n}{\sum_{p,q}}'\left\{ A_{p,q}^{+,\times}\cos\left(pl\right)\cos\left(q\lambda\right)+B_{p,q}^{+,\times}\sin\left(pl\right)\cos\left(q\lambda\right)\right.\nonumber\\&\qquad\left.+C_{p,q}^{+,\times}\cos\left(pl\right)\sin\left(q\lambda\right)+D_{p,q}^{+,\times}\sin\left(pl\right)\sin\left(q\lambda\right)\,\right\} \,,
    \label{eq:BSGKJ_residuals}
\end{align}
\label{eq:PM_Residuals}
where $\lambda=l+\gamma$ and $A_{p,q}^{+,\times}$, $B_{p,q}^{+,\times}$, $C_{p,q}^{+,\times}$, and $D_{p,q}^{+,\times}$ are polynomials of $e_t$.
The primed sum excludes the $p=q=0$ term and truncates after some finite number of harmonics $P$ depending on the order of the post-circular expansion.

~

\textbf{Method 4 (\textit{Numerical})}: 
Ref.~\cite{SusobhananGopakumar+2020} computed $s_{+,\times}(t)$ by numerically integrating $h_{+,\times}(t)$ incorporating fully 3PN-accurate orbital dynamics. 
We denote the $s_{+,\times}(t)$ computed via this method as $s_{+,\times}^\text{N}(t)$.

~

Each of these methods has its own advantages and disadvantages.
While evaluating $s_{+,\times}(t)$ given $u(t)$ and $\phi(t)$ using Method 1 is relatively inexpensive, it requires the numerical solution of the Kepler equation. 
Method 2 does not require a solution of the Kepler equation but requires expensive Bessel function evaluations while computing the Fourier coefficients. 
This Fourier series converges slowly for high eccentricities, and its truncation at some finite number of harmonics can affect the accuracy of the computed PTA signal
(see Appendix \ref{sec:pm-nharms} for an estimate of the required number of harmonics as a function of eccentricity).
Furthermore, both Method 1 and Method 2 assume a slow advance of periapsis and {may} give inaccurate results for more relativistic systems.
This is especially true of Method 2, where only the secular variations of $\omega$, given by $\gamma$, are incorporated
(see Section \ref{sec:spx_anl_accuracy} for a theoretical estimate of the error incurred in Method 1 due to this assumption).
Method 3 requires neither the solution of the Kepler equation nor the evaluation of expensive special functions but is valid only for low eccentricities.
Method 4, while being the most accurate, is also the most computationally expensive due to the application of numerical integration.


These considerations motivate us to investigate the computational cost and accuracy of the PTA signals computed using these methods, and this is what we pursue in the forthcoming sections.

\section{Computational cost of evaluating \texorpdfstring{$s_{+,\times}(t)$}{spx(t)}}
\label{sec:comp-cost}

\begin{figure}
    \centering
    \includegraphics[scale=0.1]{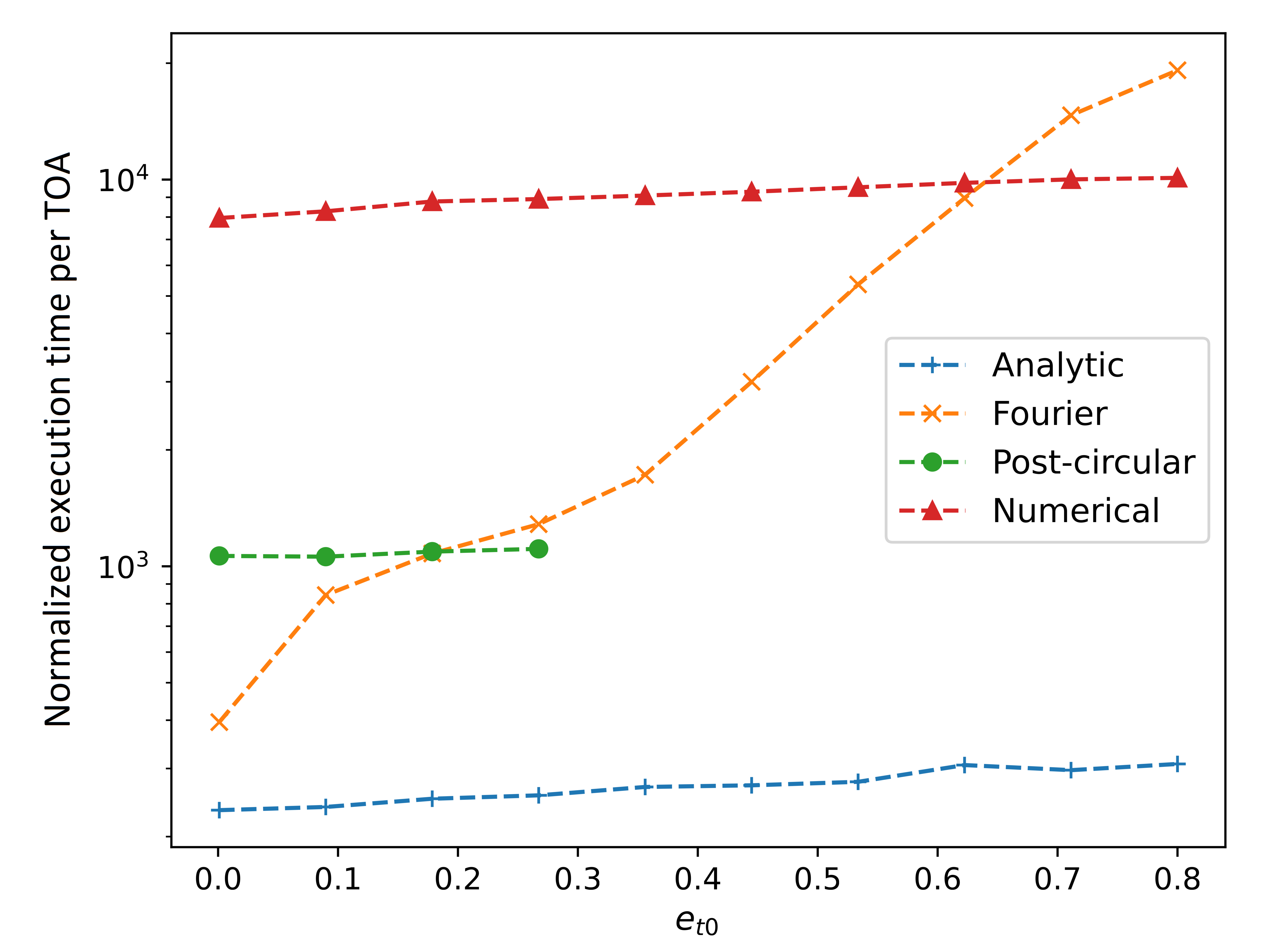}
    \caption{Computational performance comparison between different methods of computing $R(t)$ as a function of the initial eccentricity $e_{t0}$.
    Only the Earth terms are used for the comparison.
    The binary parameters used are $M=5\times 10^9 M_{\astrosun}$, $\eta=0.25$, $P_{b0}=2\pi/n_0=2$ yrs, $l_0=0$, $\gamma_0=0$, $\iota=0$, $\psi=0$.
    {The PTA signal is computed for $10^4$ TOAs spanning a time duration of 15 years (approximately 7.5 orbits).}
    The execution times are normalized using the execution time for the circular case.
    The figure shows that the analytic expressions provide $\sim50$ times improvement in performance over numerical integration, but are still $\sim$200 times slower than computing the circular PTA signals.
    The increasing trend in the computational cost for the Fourier expansion reflects the fact that the number of harmonics of the Fourier expansion required to achieve a given accuracy is an increasing function of eccentricity (see Eq.~\eqref{eq:nharms}).  
    The plot for the post-circular method is truncated because it is not valid for $e_{t0}\ge0.3$.
    }
    \label{fig:perf-comp}
\end{figure}

We now proceed to compare the cost of computing the PTA signal $R(t)$ using the approximate analytic expressions, Fourier series expansion, post-circular approximation, and numerical integration.
Figure \ref{fig:perf-comp} shows the execution time per TOA for the four methods for different eccentricities, normalized by the execution time per TOA for computing circular PTA signals.
{This includes the computation of $n$, $e_t$, $\gamma$, and $l$ using the analytic solution to equations \eqref{eq:orbit-odes} (applicable to all methods), the computation of $u$ and $\phi$ from $n$, $e_t$, $\gamma$, and $l$ (applicable to methods 1 and 4, and involves numerically solving the Kepler equation), and the evaluation of the PTA signal via an analytic expression (method 1), a truncated Fourier series (methods 2 and 3), or numerical integration (method 4).}
Figure \ref{fig:perf-comp} shows that for Analytic, Post-circular, and Numerical computations of $R(t)$, the run time does not strongly depend on the eccentricity, whereas the Fourier series expansion shows an increasing trend.
The number of harmonics required by the Fourier expansion to achieve a given degree of precision is an increasing function of eccentricity, and the trend seen in the run time for the Fourier series expansion reflects this relationship
(see Appendix \ref{sec:pm-nharms}).

The most important observation from Figure \ref{fig:perf-comp} is that the computation using the analytic expressions (Eqs.~\eqref{eq:spx_anl}) is faster than all other methods.
Specifically, it is $\sim50$ times faster than the computation using numerical integration, which is the most accurate method.
However, the fastest method for computing the eccentric PTA signals is still $\sim$200 times slower than computing the circular PTA signals.

It should be evident from Eqs.~\eqref{eq:spx_anl} and \eqref{eq:spx_pm} that Method 2 is an approximation of Method 1, which in turn is an approximation of Method 4.
Hence, it will be advantageous to use the analytic expressions to compute $s_{+,\times}(t)$ provided they are accurate enough within the region of the binary parameter space that is of interest to PTAs.
We investigate the accuracy of the analytic expressions, as compared to numerical integration, in the next section.

\section{Accuracy of the approximate analytic \texorpdfstring{$s_{+,\times}(t)$}{spx(t)} expressions}
\label{sec:spx_anl_accuracy}

While the analytic $s_{+,\times}(t)$ expressions given in Eqs.~\eqref{eq:spx_anl} are exact in the Newtonian limit, it is not so when PN effects, especially the advance of periapsis, are considered.
In the PN case, we evaluate $s_{+,\times}^\text{A}(t)$ using PN-accurate values of $u$ and $\omega$, where $\omega=\phi-f$. 
The error incurred in this prescription can be estimated as follows:
\begin{align}
\left|s_{+,\times}-s_{+,\times}^{\text{A}}\right|
&=\Bigl|s_{+,\times}-\int dt\frac{ds_{+,\times}^{\text{A}}}{dt}\Bigr|\nonumber\\
&\approx\Bigl|s_{+,\times}-\int dt\,h_{+,\times}(t)-kn\int dt\frac{\partial s_{+,\times}^{\text{A}}}{\partial\omega}\Bigr|\nonumber\\&=kn\left|\int dt\frac{\partial s_{+,\times}^{\text{A}}}{\partial\omega}\right|\,.
\end{align}
It is straightforward to show that the functions  $\mathcal{P}$ and $\mathcal{Q}$ (Eqs.~\eqref{eq:PQ_int}) have zero orbital average.
This ensures that the orbital average of $\frac{\partial{s}_{+,\times}^{\text{A}}}{\partial\omega}$ vanishes, and the error $|s_{+,\times}-s_{+,\times}^\text{A}|$ as a function of time for a given set of orbital parameters does not grow secularly with time. 
While we incur an amplitude error at the 1PN order in ${s}_{+,\times}^\text{A}(t)$ (specifically, the error is $\mathcal{O}(k)$), the phase is accurate to whichever PN order $u$ and $\phi$ are computed. 
We may expect, then, that this approximation will be valid within some low-to-moderately relativistic region of the parameter space.
To establish this, we proceed by comparing the PTA signals computed using the analytic $s_{+,\times}$ expressions (Method 1) and via numerical integration (Method 4) using a particular mismatch metric in the following subsection.

\subsection{Mismatch metric for comparing PTA signals}
\label{sec:mismatch}

In this section, we investigate the accuracy of the analytic $s_{+,\times}$ expressions against numerically integrated $s_{+,\times}$ using a mismatch metric applicable to PTA signals. 
Recall that the mismatch between two signal vectors $a$ and $b$ is usually defined as 
\begin{equation}
\text{mismatch}[a,b]=1-\frac{(a,b)}{\sqrt{(a,a)(b,b)}}\,,    
\end{equation}
where $(a,b)$ represents an appropriately defined inner product that takes into account the detector response. 
In the case of a PTA dataset, $a$ and $b$ will be some signals evaluated at each TOA, and the appropriate notion of the inner product is given by Ref.~\cite{vanHaasterenLevin2010} as 
\footnote{{Note that the inner product $(a,b)$ is slightly different from the inner product $\left\langle a,b\right\rangle$ defined in Ref.~\cite{vanHaasterenLevin2010}. 
Specifically, $(a,b)=\left\langle a,Rb\right\rangle$ in their notation where $R$ is a projection operator defined in Eq. (16) of Ref.~\cite{vanHaasterenLevin2010}.}}
\begin{equation}
(a,b)=a^{T}Kb\,,   
\label{eq:inner-prod}
\end{equation}
where 
\begin{equation}
K=C^{-1}-C^{-1}\mathfrak{M}(\mathfrak{M}^{T}C^{-1}\mathfrak{M})^{-1}\mathfrak{M}^{T}C^{-1}\,,
\end{equation}
$\mathfrak{M}$ is the $N\times p$ dimensional pulsar timing design matrix {(not to be confused with the total mass $M$)},
$C$ is the $N\times N$ dimensional TOA covariance matrix, 
$p$ is the number of timing model parameters, and 
$N$ is the number of TOAs.
For simplicity, we assume an ideal pulsar dataset with no correlated noise and equal TOA measurement uncertainties $\sigma$. 
In this case, $C=\sigma^{2}I$, where $I$ is the $N\times N$ dimensional identity matrix, and $K$ simplifies to
\begin{equation*}
    K=\sigma^{-2}\left(I-\mathfrak{M}(\mathfrak{M}^{T}\mathfrak{M})^{-1}\mathfrak{M}^{T}\right)\,,
\end{equation*}
and we note that $\text{mismatch}[a,b]$ is independent of $\sigma$.
Therefore, we set $\sigma=1$ and write 
\begin{equation}
K=I-\mathfrak{M}(\mathfrak{M}^{T}\mathfrak{M})^{-1}\mathfrak{M}^{T}\,.
\end{equation}
The most important contributions to $\mathfrak{M}$ arise from fitting TOAs for the pulse phase offset, pulsar spin frequency, spin frequency derivative, and the sky location of the pulsar. 
The effects of fitting these parameters are, approximately, to remove a constant offset, a straight line, a quadratic function, and a sinusoid of period 1 year respectively
{(i.e., a function of the form $a_0+a_1 (t-t_m)+a_2 (t-t_m)^2 +a_c \cos\frac{t}{1\text{ yr}} + a_s \sin\frac{t}{1\text{ yr}}$ for some constant coefficients $a_i$ and reference time $t_m$)}.
Hence, given some TOAs $t_{1}\,...\,t_{N}$, the design matrix $M$
can be approximated as
\[
\mathfrak{M}=\begin{bmatrix}1 & 1 & ... & 1\\
t_{1}/t_{m}-1 & t_{2}/t_{m}-1 & ... & t_{N}/t_{m}-1\\
\left(t_{1}/t_{m}-1\right)^{2} & \left(t_{2}/t_{m}-1\right)^{2} & ... & \left(t_{N}/t_{m}-1\right)^{2}\\
\sin\left(\frac{2\pi t_{1}}{\text{1 yr}}\right) & \sin\left(\frac{2\pi t_{2}}{\text{1 yr}}\right) & ... & \sin\left(\frac{2\pi t_{N}}{\text{1 yr}}\right)\\
\cos\left(\frac{2\pi t_{1}}{\text{1 yr}}\right) & \cos\left(\frac{2\pi t_{2}}{\text{1 yr}}\right) & ... & \cos\left(\frac{2\pi t_{N}}{\text{1 yr}}\right)
\end{bmatrix}\,,
\]
where $t_{m}$ is the median TOA. 
Note that this idealized scenario corresponds to a highly sensitive PTA dataset and will therefore produce worse mismatch values for a given pair of signals as compared to a realistic dataset. 
Hence, this is a more stringent comparison between signals than one done using a realistic dataset.

\begin{figure}[t]
\begin{centering}
\includegraphics[scale=0.6]{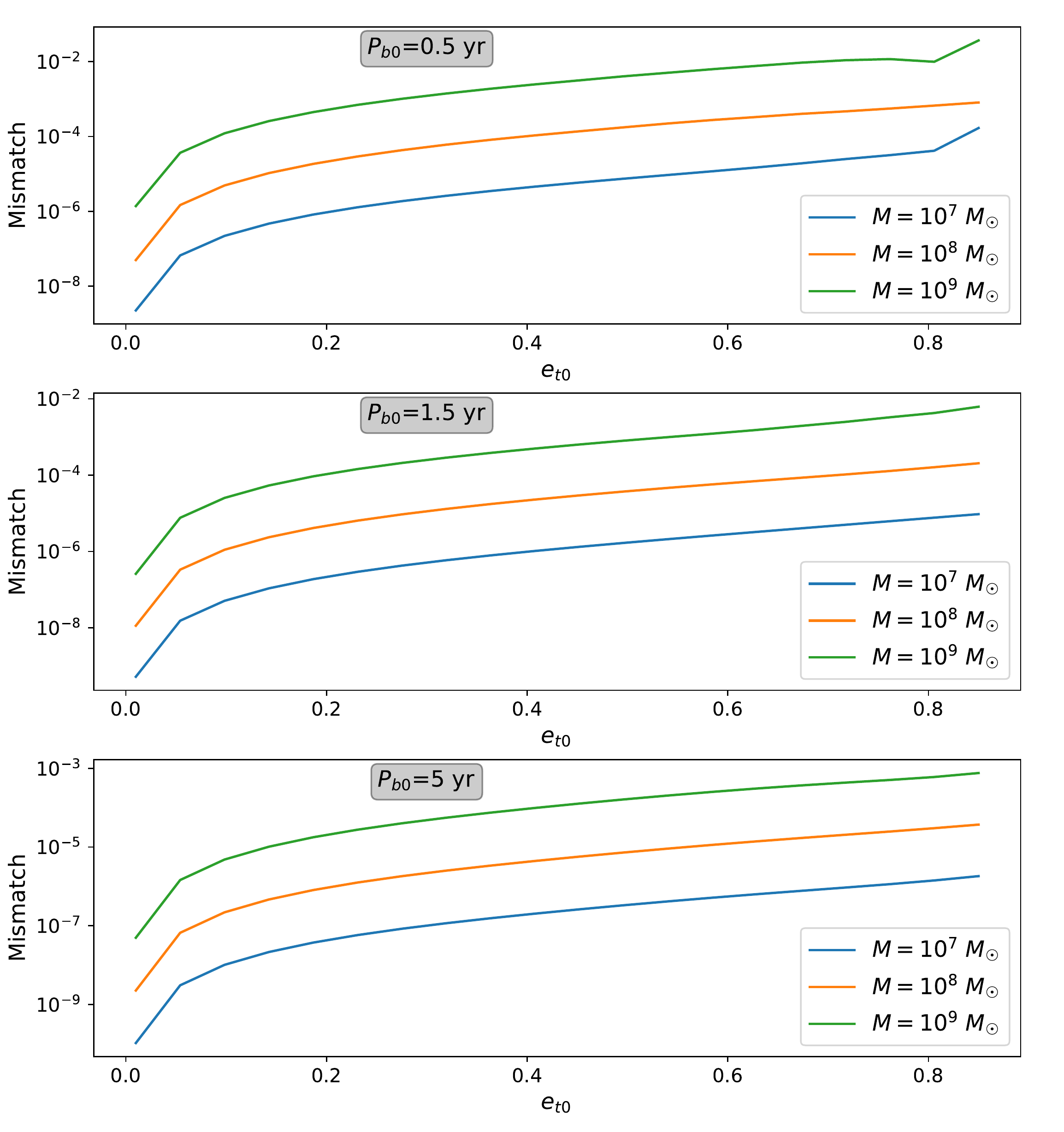}
\end{centering}
\caption{Mismatch between approximate analytic PTA signals and numerically integrated PTA signals for different values of $M$, $P_{b0}$, and $e_{t0}$.
The plots correspond to $\text{RA}_{\text{psr}}=4^{h}37^{m}15.81476^{s}$,
$\text{DEC}_{\text{psr}}=-47^{\circ}15'8.6242"$, $\text{RA}_{\text{gw}}=4^{h}0^{m}0^{s}$,
$\text{DEC}_{\text{gw}}=-45^{\circ}0'0"$, $\eta=0.25$, $\psi=0$,
$\iota=0$, $l_{0}=0$, $\gamma_{0}=0$.
Only the Earth terms are compared.
The mismatch is only $\sim0.01$ even for $M=10^9M_{\astrosun}$, $e_{t0}=0.85$ and $P_{b0}=0.5$ year.}
\label{fig:mismatch}
\end{figure}

We now plot in Figure \ref{fig:mismatch} the mismatch between approximate analytic PTA signals given by Eqs.~\eqref{eq:spx_anl} and numerically integrated PTA signals for different values of $M$, $P_{b0}=2\pi/n$ and $e_{t0}$. 
We fix the other binary parameters and the pulsar and GW source sky locations since the largest contribution to the mismatch arises from the non-exact treatment of the advance of periapsis while doing the integrals in Eqs.~\eqref{eq:PQ_int}, and the advance of periapsis per orbit $k$ only depends on $M$, $n$ and $e_t$ at the leading order \cite{DamourDeruelle1985}.

\begin{figure*}
\begin{centering}
\makebox[\textwidth][c]{
    \includegraphics[scale=0.6]{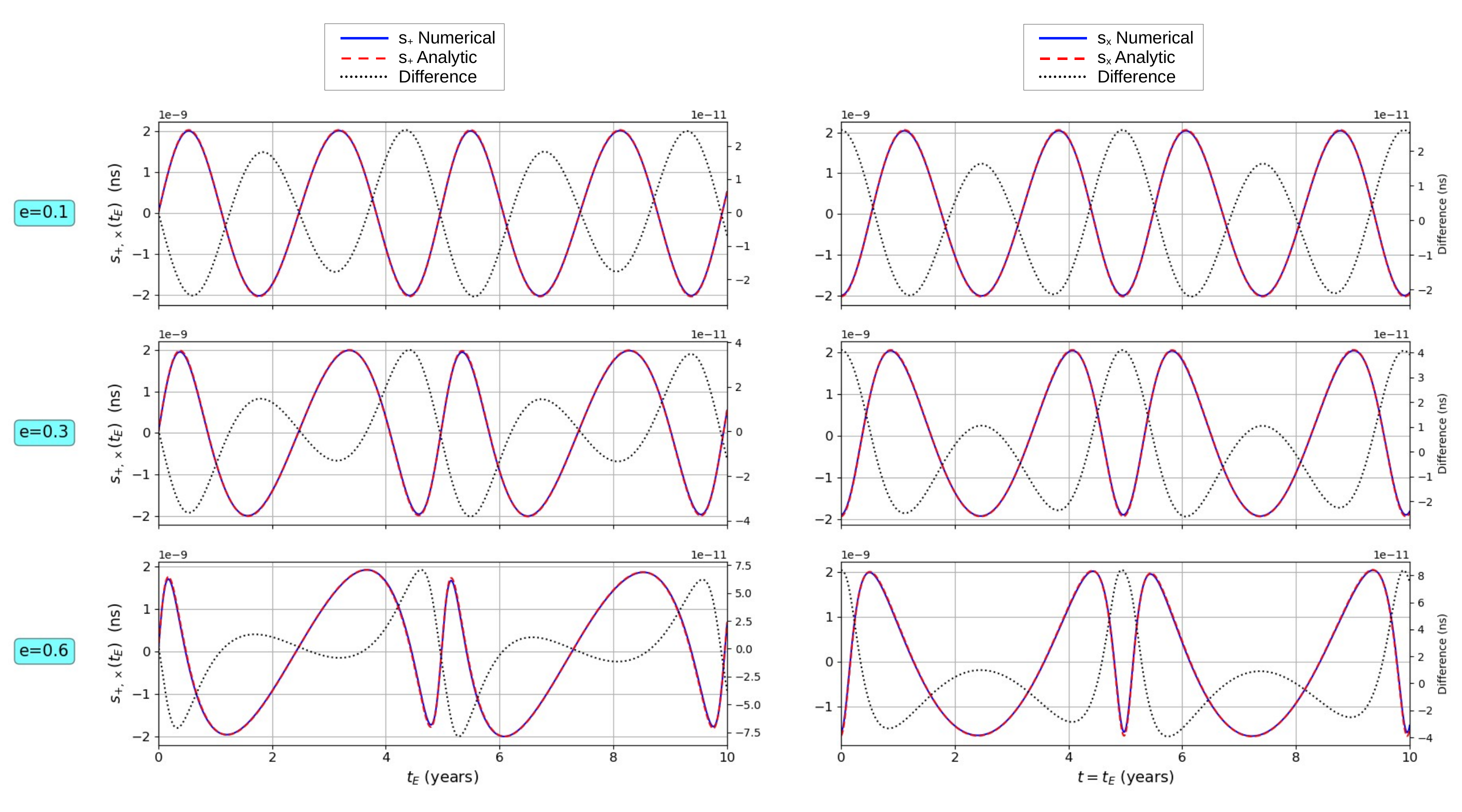}
}
\par\end{centering}
\caption{Comparison between the analytic PTA signal $s_{+,\times}^{\text{A}}(t_E)$ (red dashed line) and the numerically integrated PTA signal $s_{+,\times}^{\text{N}}(t_E)$ (blue solid line) for different eccentricities, $M=10^{9}M_{\astrosun}$, $\eta=0.25,$ $P_{b}=5$ years, $\psi=0$, $\iota=0$, $l_{0}=0$, and $\gamma_{0}=0$.
The difference $s_{+,\times}^{\text{A}}(t_E)-s_{+,\times}^{\text{N}}(t_E)$ is plotted using black dotted lines with a separate Y axis on the right-hand side of each plot.
Only the Earth-term contributions are shown. 
The left panels show $s_{+}$ and the right panels show $s_{\times}$.
The difference between the curves is visually discernible from the red and blue curves only for $e=0.6$  where the curves show sharp features.
}
\label{fig:spx-comparison}
\end{figure*}

In Figure \ref{fig:mismatch}, the mismatch grows {larger} for increasing $M$ and $e_{t0}$ and decreasing $P_{b}$ as expected. 
Interestingly, even for high eccentricities ($e>0.8$), the mismatch is only $\sim0.01$, indicating the good quality of the approximation. 
Hence, we may conclude that this approximation can be used for masses up to $10^{9}M_{\astrosun}$, eccentricities up to $0.85$, and orbital periods as small as $0.5$ years. 
This can also be seen in Figure \ref{fig:spx-comparison}, which shows a visual comparison of the approximate analytic and numerically integrated waveforms.

\section{Efficient computation of  \texorpdfstring{$R(t)$}{R(t)} for realistic PTA datasets}
\label{sec:spline-method}

The relatively efficient method discussed above for computing $R(t)$ still requires costly  hypergeometric function evaluations to solve the orbital evolution analytically (see Eqs.~30 of Ref.~\cite{SusobhananGopakumar+2020}) as well as the inversion of the Kepler equation (Eq.~\eqref{eq:kepler}).
Furthermore, it is in general not possible to obtain a fully analytic prescription for computing $R(t)$ when higher order reactive PN effects \cite[e.g.][]{DamourGopakumarIyer2004} are taken into account.
In this section, we further reduce the computational cost of evaluating $R(t)$ given a set of TOAs by leveraging the properties of realistic PTA datasets.

PTA experiments measure TOAs from their radio observations via two different techniques, namely the traditional narrowband technique \citep[e.g.][]{AlamArzoumanian+2020} and the modern wideband technique \citep[e.g.][]{AlamArzoumanian+2020b}, resulting in two types of datasets.
These techniques differ in how they model the observing frequency-dependent effects in the pulsar signal, such as the interstellar dispersion and the frequency-dependent evolution of the integrated pulse profile.
The narrowband technique measures multiple TOAs per observation epoch by splitting broadband observations into multiple frequency sub-bands, such that the frequency-dependent effects are reduced in each sub-band.
The wideband technique employs principal component decomposition of the frequency-resolved pulse profiles to model the interstellar dispersion and the frequency-dependent profile evolution, resulting in one TOA and one DM measurement per observation \cite{PennucciDemorestRansom2014,Pennucci2019}.
Further, the PTA datasets usually have observational cadences of the order of weeks, with time spans up to decades.

In narrowband datasets, the sub-banded TOAs derived from the same observation are typically only a fraction of a second apart, and therefore the $R(t)$ values evaluated at those TOAs will not show any appreciable variation.
There can be multiple TOAs per {epoch} even in the case of wideband datasets, e.g., in the case of simultaneous multi-band observations \cite{JoshiGopakumar+2022}.
In such cases, it is not necessary to evaluate $R(t)$ independently at each TOA.
Rather, one can evaluate $R(t)$ only at unique epochs $t_i$ and interpolate the $R(t)$ values for the TOAs $t_{ij}$ at that epoch as
\[
R(t_{ij}) = R(t_i) + h(t_i)(t_i-t_{ij})\,.
\]
Note that this computation requires $h(t_i)$ at each epoch, which is cheap to compute together with $R(t_i)$.
The correction $h(t_i)(t_i-t_{ij})$ has a magnitude that is $n|t_i-t_{ij}|$ times less than $R(t)$, which is only $\sim10^{-2}$ for $n\sim 100$ nHz and $|t_i-t_{ij}|\sim 1$ day.
The relative error incurred in this approximation is of the order of $(n|t_i-t_{ij}|)^2\sim10^{-4}$ for $n\sim 100$ nHz and $|t_i-t_{ij}|\sim 1$ day.
Therefore, $R(t)$ has to be computed only once every day, even if the TOAs on that day were not obtained from simultaneous observations.
If required, the approximation can be improved by including the $\mathcal{O}((n|t_i-t_{ij}|)^2)$ term, which requires the computation of $h'(t)=dh/dt$.
$h'(t)$ can be derived easily by differentiating Eq.~\eqref{eq:hpx} and is cheap to compute together with $R(t)$ like $h(t)$.
In practice, we implement this by interpolating the $R(t)$ values computed for every observation epoch using a cubic Hermite spline \cite{Kreyszig2005} to compute $R(t)$ at every TOA.

\begin{figure*}
    \centering
    \includegraphics[scale=0.8]{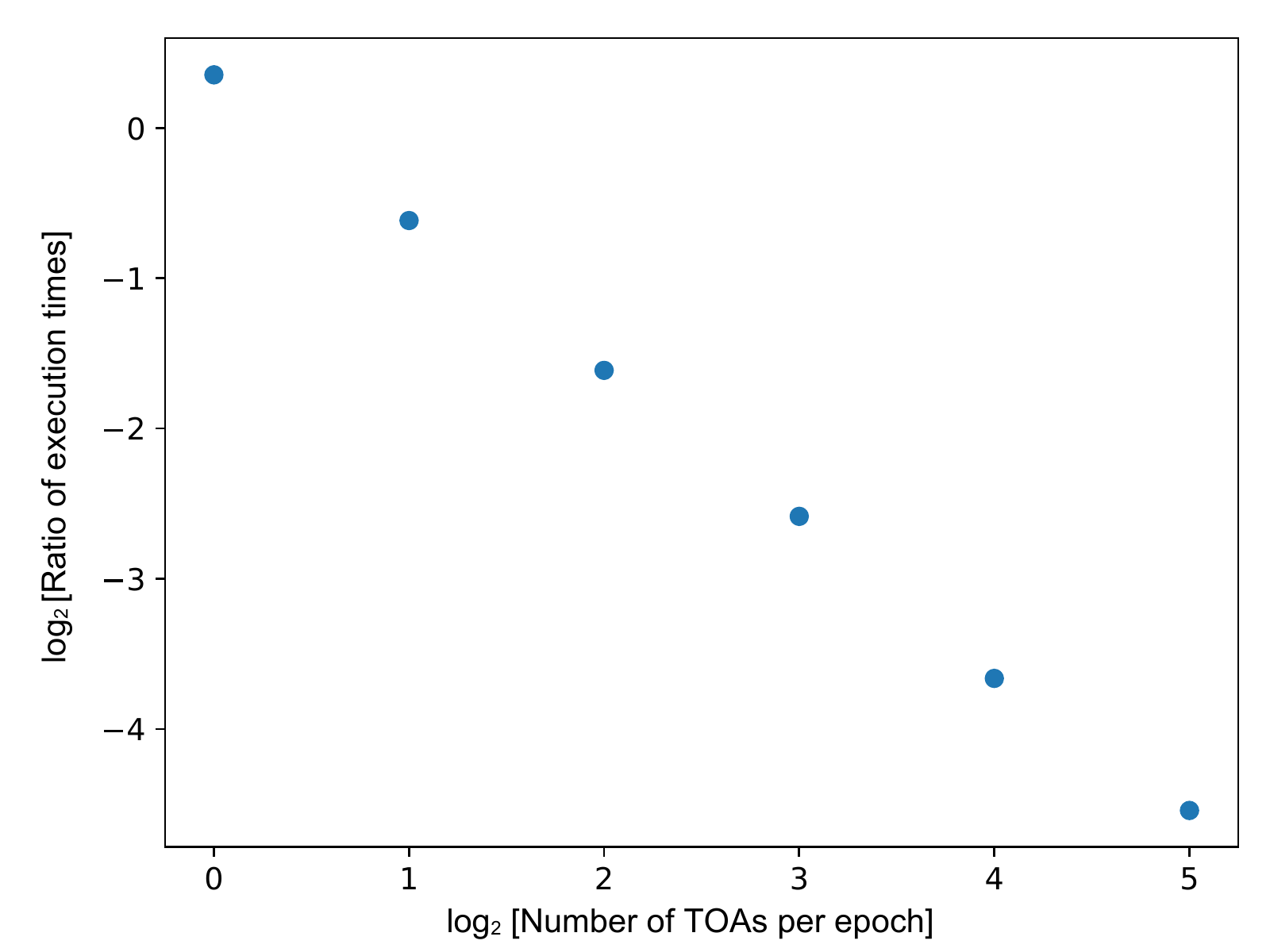}
    \caption{Ratio between execution times of computing $R(t)$ at every TOA using the interpolation method  vs independently at each TOA. 
    }
    \label{fig:interp-perf}
\end{figure*}

This method can, in an ideal scenario, reduce the computational cost of  evaluating $R(t)$ by a factor equal to the average number of TOAs per epoch, without any appreciable loss of accuracy. 
For example, the NANOGrav 12.5-year data release has  $\sim$58 TOAs per epoch for the narrowband dataset and $\sim$1.4 TOAs per epoch for the wideband dataset on average for PSR J1909$-$3744 \cite{AlamArzoumanian+2020,AlamArzoumanian+2020b}.
This method can be applied to TOAs of multiple pulsars if only the Earth term is considered, providing a further performance improvement of a factor that is proportional to the number of pulsars in the best-case scenario.
The performance gained by employing this method for different numbers of TOAs per epoch is plotted in Figure \ref{fig:interp-perf}.
This figure clearly shows the expected decreasing trend in the ratio between the execution times of computing $R(t)$ using the interpolation method  and computing $R(t)$  independently at each TOA.


\section{The \texorpdfstring{\texttt{GW\lowercase{ecc.jl}}}{GWecc} package}
\label{sec:gwecc.jl}

We have implemented the optimal method for computing the eccentric SMBHB PTA signal, including the analytic solution to the quadrupolar order orbital evolution (section \ref{sec:phasing}), the approximate analytical expression for the PTA signal (Method 1 in section \ref{sec:PTA-signal-compute}), and the spline-based method for accelerating the PTA signal computation for realistic datasets (section \ref{sec:spline-method}), in a package named \GWeccjl{}\footnote{Available at \url{https://github.com/abhisrkckl/GWecc.jl}.}.
This package re-implements parts of the \GWecc{} C++ package \footnote{The \GWecc{} package (\url{https://github.com/abhisrkckl/GWecc}) implements the PTA signal computation using Methods 1-4.} \cite{SusobhananGopakumar+2020,Susobhanan2020} in Julia for ease of Python integration and unit testing.
A Python interface (\texttt{enterprise\_gwecc}) is provided for the sake of easy usage with \texttt{ENTERPRISE}, and includes the function \texttt{eccentric\_pta\_signal} for computing the PTA signal given sets of SMBHB parameters and TOAs, and the function \texttt{gwecc\_block} which creates an \enterprise{} deterministic signal  (\texttt{enterprise.signals.deterministic\_signals.Deterministic}) object.
Similar functions \texttt{eccentric\_pta\_signal\_1psr} and \texttt{gwecc\_1psr\_block} are also provided for the simplified single-pulsar PTA signal (see section \ref{sec:pta-signal-1psr} below). 
This package was thoroughly unit and integration tested to ensure the correctness of the generated PTA signals.

\section{Single-pulsar search for eccentric SMBHBs using NANOGrav 12.5-year data of PSR J1909--3744}
\label{sec:1psr-search}

We now proceed to demonstrate our methods by searching for and constraining eccentric SMBHB signals using the NANOGrav 12.5-year narrowband data \cite{AlamArzoumanian+2020} of PSR J1909$-$3744 using a Bayesian framework.

\subsection{Simplified PTA signal expression for the single-pulsar case}
\label{sec:pta-signal-1psr}

We begin our analysis by investigating the parameter degeneracies in the single-pulsar case and simplifying the PTA signal expression \eqref{eq:pta-res} for this case.
For this purpose, we rewrite equations \eqref{eq:pta-res} and \eqref{eq:spx_anl} in the form
\begin{align}
R(t)=\begin{bmatrix}a_{0} & a_{1} & a_{2}\end{bmatrix}\begin{bmatrix}\mathcal{A}_{0}(t)-\mathcal{A}_{0}(t-\Delta_{p})\\
\mathcal{A}_{1}(t)-\mathcal{A}_{1}(t-\Delta_{p})\\
\mathcal{A}_{2}(t)-\mathcal{A}_{2}(t-\Delta_{p})
\end{bmatrix}\,,
\label{eq:R(t)_mat}
\end{align}
where we have defined the projection coefficients
\begin{align}
\begin{bmatrix}a_{0} & a_{1} & a_{2}\end{bmatrix}=S_{0}\begin{bmatrix}F_{+} & F_{\times}\end{bmatrix}\begin{bmatrix}\cos2\psi & -\sin2\psi\\
\sin2\psi & \cos2\psi
\end{bmatrix}\begin{bmatrix}s_{i}^{2} & \left(c_{\iota}^{2}+1\right) & 0\\
0 & 0 & 2c_{\iota}
\end{bmatrix}\begin{bmatrix}1 & 0 & 0\\
0 & \cos2\gamma_{0} & -\sin2\gamma_{0}\\
0 & \sin2\gamma_{0} & \cos2\gamma_{0}
\end{bmatrix}\,,
\end{align}
and the signal components
\begin{align}
   \begin{bmatrix}\mathcal{A}_{0}\\
\mathcal{A}_{1}\\
\mathcal{A}_{2}
\end{bmatrix}=\varsigma \begin{bmatrix}1 & 0 & 0\\
0 & \cos2\varpi & -\sin2\varpi\\
0 & \sin2\varpi & \cos2\varpi
\end{bmatrix}\begin{bmatrix}\mathcal{R}\\
\mathcal{Q}\\
\mathcal{P}
\end{bmatrix}\,,
\label{eq:A_funcs}
\end{align}
such that $S=H/n$, $S_0$ is $S$ evaluated at $t_0$, $\varsigma=S/S_0$, and $\varpi=\omega-\gamma_0$.
Since we are considering only one pulsar, the antenna pattern vector $[F_+,F_\times]$ can be absorbed into a magnitude $F=\sqrt{F_+^2+F_\times^2}$ and a redefinition of the polarization angle $\psi'=\psi-\frac{1}{2}\arctan\frac{F_\times}{F_+}$. 
This leads to
\begin{align}
    \begin{bmatrix}a_{0} & a_{1} & a_{2}\end{bmatrix}=S_{0}F\begin{bmatrix}\beta_{0} & \beta_{1} & \beta_{2}\end{bmatrix}\,,
\end{align}
where
\begin{align}
    \begin{bmatrix}\beta_{0}&
\beta_{1}&
\beta_{2}
\end{bmatrix}&=
\begin{bmatrix}1 & 0\end{bmatrix}\begin{bmatrix}\cos2\psi' & -\sin2\psi'\\
\sin2\psi' & \cos2\psi'
\end{bmatrix}\begin{bmatrix}s_{i}^{2} & \left(c_{i}^{2}+1\right) & 0\\
0 & 0 & 2c_{i}
\end{bmatrix}\begin{bmatrix}1 & 0 & 0\\
0 & \cos2\gamma_{0} & -\sin2\gamma_{0}\\
0 & \sin2\gamma_{0} & \cos2\gamma_{0}
\end{bmatrix}\nonumber\\
&=\begin{bmatrix}\left(1-c_{\iota}^{2}\right)\cos(2\psi')\\
\left(1+c_{\iota}^{2}\right)\cos(2\gamma_{0})\cos(2\psi')-2c_{\iota}\sin(\ensuremath{2\gamma}_{0})\sin(2\psi')\\
-\left(1+c_{\iota}^{2}\right)\sin(2\gamma_{0})\cos(2\psi')-2c_{\iota}\cos(2\gamma_{0})\sin(2\psi')
\end{bmatrix}^T\,.
\end{align}
We now define the spherical polar representation $(\beta, \sigma,\rho)$ of the vector $[\beta_0,\beta_1,\beta_2]$ such that $\beta=\sqrt{\beta_0^2+\beta_1^2+\beta_1^2}$, $\tan\sigma=\sqrt{\beta_1^2+\beta_2^2}/\beta_0$, $\tan\rho=\beta_2/\beta_1$.
This allows us to write
\begin{align}
    \begin{bmatrix}a_{1} & a_{2} & a_{3}\end{bmatrix}=S_{0}F\beta\begin{bmatrix}\cos\sigma & \sin\sigma\cos\rho & \sin\sigma\sin\rho\end{bmatrix}\,.
    \label{eq:a_coeffs}
\end{align}
Defining the effective amplitude $\zeta_0=S_0 F \beta$, the PTA signal can thus be written in the simplified form \begin{align}
R(t)=&\zeta_{0}\left[\cos\sigma\left(\mathcal{A}_{0}(t)-\mathcal{A}_{0}(t-\Delta_{p})\right)+\sin\sigma\cos\rho\left(\mathcal{A}_{1}(t)-\mathcal{A}_{1}(t-\Delta_{p})\right)\right.
\nonumber\\
&+\left.\sin\sigma\sin\rho\left(\mathcal{A}_{2}(t)-\mathcal{A}_{2}(t-\Delta_{p})\right)\right]\,.
\label{eq:R(t)_1psr}
\end{align}
This simplified form manifestly shows the insensitivity of a single-pulsar search for separately detecting the two GW polarizations and localizing the GW source. 

\begin{figure*}
    \centering
    \makebox[\textwidth][c]{
    \includegraphics[scale=0.75]{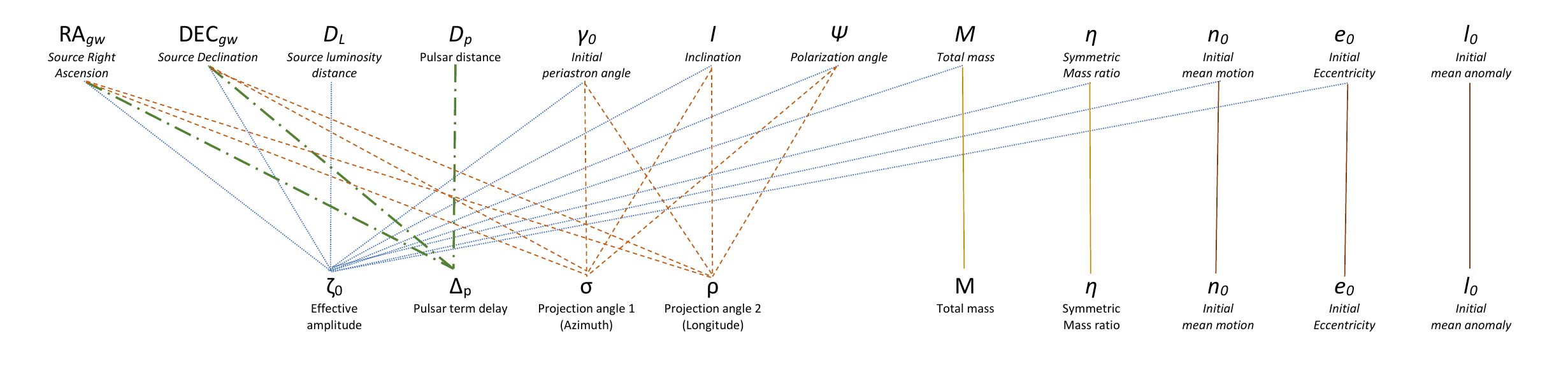}}
    \caption{Relationships between the reduced parametrization for the single-pulsar (bottom row) and the general parametrization (top row). 
    The different colors and line styles are used only to enhance visibility.}
    \label{fig:param-graph}
\end{figure*}

We list the free parameters of the PTA signal in the single-pulsar case for the forthcoming analysis: 
the projection parameters $(\zeta_0, \sigma, \rho)$ and the shape parameters $(M, \eta, n_0, e_0, l_0, \Delta_p)$.
We treat $\Delta_p$ as a free parameter because we had absorbed the antenna pattern (and hence, the GW source coordinates) into the three projection parameters.
The general expression \eqref{eq:R(t)_px} for the PTA signal has one pulsar-dependent parameter $D_p$ and eleven common parameters $(\text{RA}_\text{gw}, \text{DEC}_\text{gw}, D_L, M,\eta,n_0,e_0,l_0,\gamma_0,\iota,\psi)$.
On the other hand, the simplified expression \eqref{eq:R(t)_1psr} derived above only has nine parameters in total; 
i.e., we have removed the parameter degeneracies arising in the single pulsar case.
The relationships between the original parametrization and the reduced parametrization in the single-pulsar case are illustrated in Figure \ref{fig:param-graph}.

We note in passing that similar degeneracies may arise in the cases of PTAs containing few (two or three) pulsars, and such cases will be interesting to explore in future work.

\subsection{Data analysis}

The NANOGrav 12.5-year narrowband dataset of PSR J1909$-$3744 consists of 22633 TOAs obtained using the Green Bank Telescope at 820 MHz and 1.4 GHz over a time span of 12.7 years from 2004 to 2017, and the corresponding timing model \cite{AlamArzoumanian+2020}.
Each TOA is transformed from its observatory frame to the SSB frame using the DE436 solar system ephemeris.
The dispersion measure (DM) variations present in the TOAs are corrected for by applying the DMX parameters, which provide a piecewise-constant model for DM variations.

The pulsar timing residuals $\delta t$ can be expressed as the sum of different components:
\begin{equation}
    \delta t = \mathfrak{M}\boldsymbol{\epsilon} + n_w + n_r + R\,,
\end{equation}
where we recall that $\mathfrak{M}$ is the pulsar timing design matrix, and $\boldsymbol\epsilon$ is a vector representing small deviations from the best-fit timing model parameter values that arise while incorporating signal components unaccounted for in the timing model.
$n_w$ and $n_r$ represent the time-uncorrelated (white) and time-correlated (red) noise, and $R$ represents the GW signal emitted by an individual SMBH source given by equation \eqref{eq:R(t)_1psr} for single-pulsar analysis.

Following NANOGrav analyses such as Ref. \cite{ArzoumanianBaker+2023}, we model the white noise $n_w$ using three parameters for each telescope backend: EFAC, EQUAD, and ECORR.
EFAC scales the TOA uncertainties by a multiplicative factor, EQUAD adds to the TOA uncertainties in quadrature, and ECORR describes the correlation between TOAs derived from the same observation.
Recent studies have shown that the red noise $n_r$ is the sum of a common process $n_{rc}$ and a pulsar-dependent component $n_{rp}$  \cite[e.g.][]{ArzoumanianBaker+2020b}.
However, since we are using only one pulsar in our analysis, it is not possible to disentangle $n_{rc}$ and $n_{rp}$, and we model $n_{r}$ as a single Gaussian red noise process with a power-law spectrum.
The red noise power-law spectrum is divided into 30 linearly spaced frequency bins ranging from $1/T_\text{obs}$ to $30/T_\text{obs}$, where $T_\text{obs}$ is the total observation time span of the pulsar.

The prior distributions for the eccentric SMBHB signal parameters we use in our analysis are listed in Table \ref{tab:priors}.
We use two types of priors for the amplitude $\zeta_0$: the detection analysis is performed using a uniform prior on $\log_{10}\zeta_0$, whereas the upper limits are computed using a uniform prior on $\zeta_0$ (`LinearExp' in Table \ref{tab:priors}).
Note that uniform priors on $\sigma$ and $\rho$ do not correspond to uniform priors on $\psi$, $\cos\iota$, and $\gamma_0$, and are used here only for simplicity.
A fixed value of $D_p$ and a uniform distribution on the GW source location  corresponds approximately to a uniform distribution on $\Delta_p$ between 0 and $D_p/c$.
To account for the measurement uncertainty in $D_p$, we have set the upper bound of the $\Delta_p$ prior distribution to be at $(D_p+\sigma_{D_p})/c$, where $\sigma_{D_p}$ is the $1\sigma$ uncertainty of the $D_p$ measurement.
The prior upper bounds on the GW frequency in the circular limit $f_{gw}=n_0/\pi$, $e_0$, and $M$ are chosen to exclude regions of the parameter space where the post-Keplerian description breaks down.
Although $\eta\in(0,0.25]$ and $e\in[0,1)$ by definition, the lower limits of $\eta$ and $e$ are set at small non-zero numbers to avoid division by zero errors.
The fiducial time $t_0$ is fixed at the latest TOA in the dataset so that no merger occurs within the data span for any given set of parameters.

\begin{table}[]
    \centering
    \begin{tabular}{c|c|c}
     \hline
\textbf{Parameter} & \textbf{Unit}  & \textbf{Prior}         \\ \hline
$\log_{10}\zeta_0$  & s               & Uniform[-11, -5] /  LinearExp[-11, -5]  \\ 
$\sigma$           & rad             & Uniform[0, $\pi$]  \\ 
$\rho$             & rad             & Uniform[0, $2\pi$] \\ 
$\log_{10}M$       & $M_{\astrosun}$ & Uniform[6, 9]      \\ 
$\eta$             &                 & Uniform[0.01, 0.25]   \\ 
$\log_{10}f_{gw}$  & Hz              & Uniform[-9, -7]    \\ 
$e_0$              &                 & Uniform[0.01, 0.8] \\ 
$l_0$              & rad             & Uniform[0, $2\pi$] \\ 
$\Delta_p$         & yr              & Uniform[0, $(D_p+\sigma_{D_p})/c$]  \\ \hline
    \end{tabular}
    \caption{Prior distributions for the PTA signal parameters. $f_\text{gw}=n_0/\pi$ is the gravitational wave frequency in the circular limit at the fiducial time $t_0$.
    $t_0$ is fixed to be at the latest TOA.}
    \label{tab:priors}
\end{table}

We utilize Bayesian methods to obtain the posterior distribution of the model parameters.
We use the \GWeccjl{} package to evaluate the PTA signal given by Eq. \eqref{eq:R(t)_1psr}, the \enterprise{} package to evaluate the corresponding prior distribution and the likelihood function (given in, e.g., Ref \cite{ArzoumanianBrazier+2014}), and the \texttt{PTMCMCSampler} package \cite{EllisvanHaasteren2017} to draw samples from the posterior distribution.\footnote{\texttt{PTMCMCSampler} implements the Parallel Tempering Markov Chain Monte Carlo algorithm \cite{Sambridge2013}, and allows us to use different types of MCMC proposals, including user-defined ones with tunable weights.
We use the following types of proposals in combination with Parallel Tempering swaps: Single Component Adaptive Metropolis (SCAM), Adaptive Metropolis (AM), Differential Evolution (DE), draws from the prior distribution, and draws from an empirical distribution for pulsar red noise parameters generated from a noise-only analysis.
The latter two proposals are implemented using the \texttt{enterprise\_extensions} package \cite{TaylorBaker+2021}.
See, e.g., Refs \cite{Ellis2013,ArzoumanianBrazier+2014} and references therein for more details.}
The search scripts used in this work are available at \url{https://github.com/abhisrkckl/GWecc\_1psr\_search}.

\subsection{Results}

\begin{figure}[ht]
    \centering
    \includegraphics[scale=0.83]{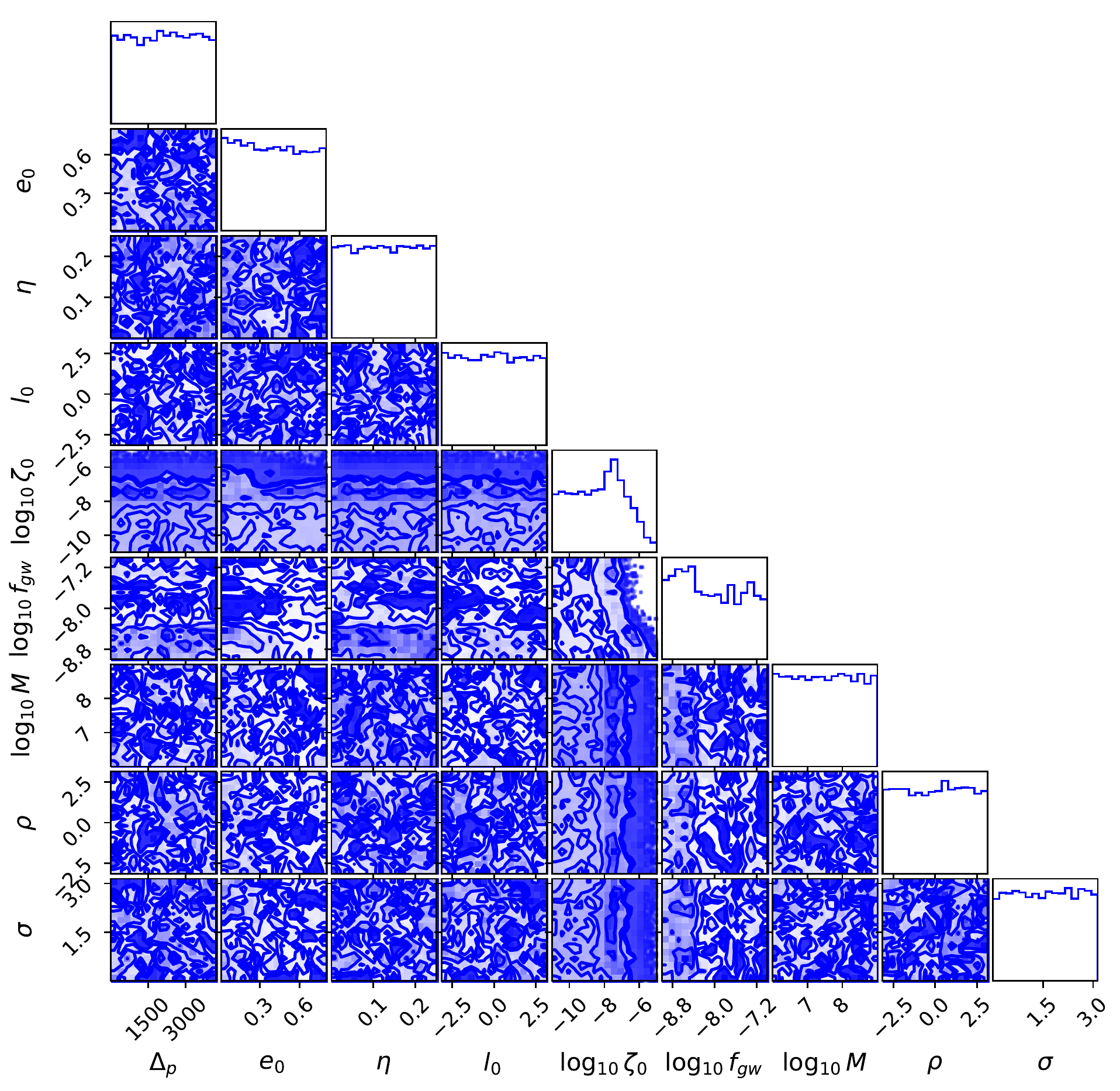}
    \caption{Posterior distribution of the continuous GW parameters. The units of each parameter are listed in table \ref{tab:priors}.
    The plot does not show any evidence for the detection of an eccentric SMBHB GW signal.}
    \label{fig:gwecc-posterior}
\end{figure}

The posterior distribution of the continuous GW parameters marginalized over the noise parameters,  obtained from a detection analysis (i.e., using a uniform prior on $\log_{10}\zeta_0$), is shown in Fig. \ref{fig:gwecc-posterior} as a corner plot.
Fig. \ref{fig:gwecc-posterior} does not show any evidence for an individual eccentric SMBHB GW signal, consistent with previous studies. 
The Bayes factor between the models with and without the eccentric individual SMBHB GW signal, estimated using the Savage-Dickey formula, is approximately $0.9$, and does not favor the detection of a GW signal (see, e.g., \cite{AggarwalArzoumanian+2019} for details on the computation of the Savage-Dickey Bayes factor). 

\begin{figure}[ht]
    \centering    \includegraphics[scale=0.65]{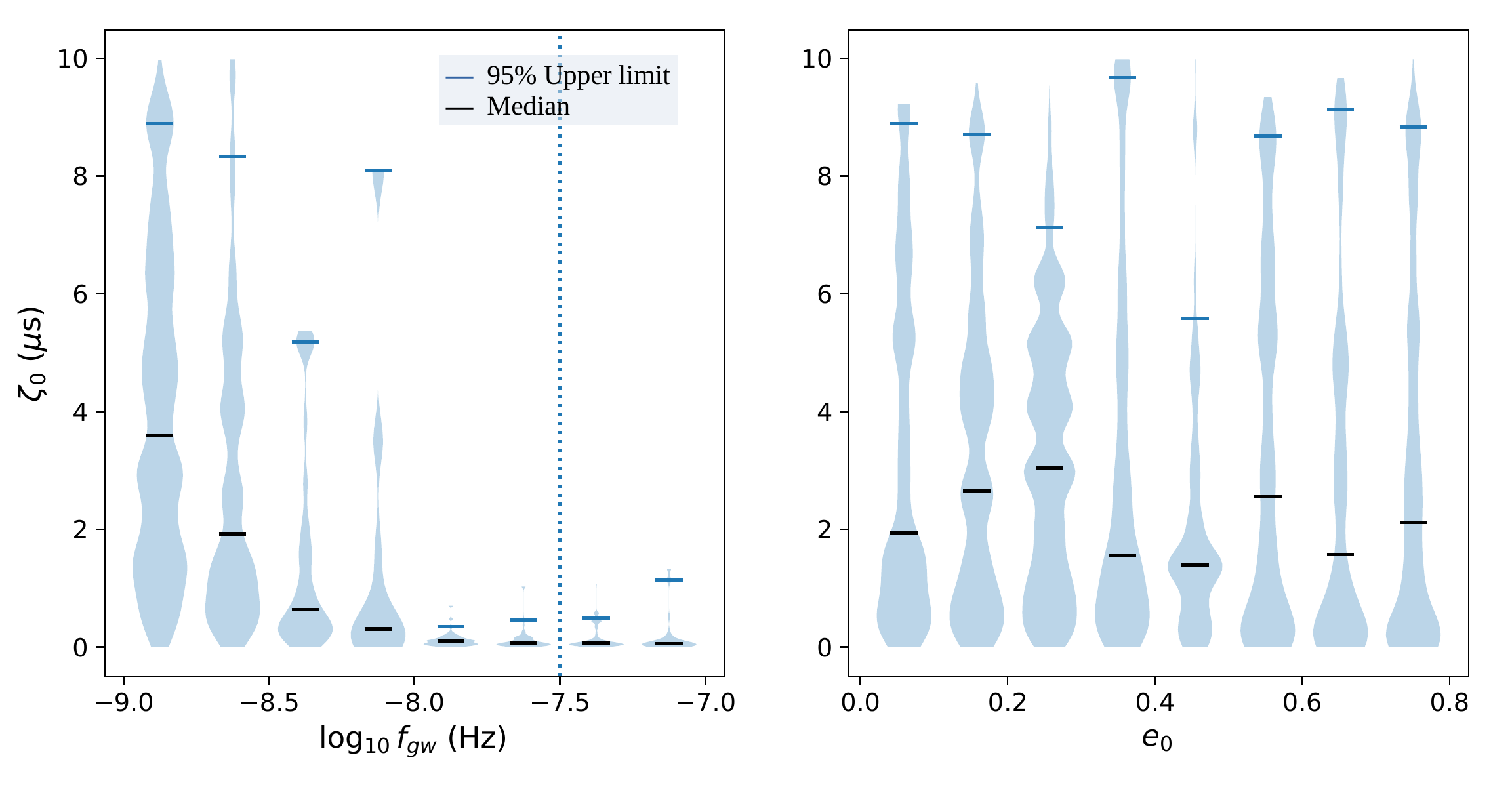}
    \caption{Violin plots of  $\zeta_{0}$, binned in  $\log_{10}f_{gw}$ (left plot) and $e_0$ (right plot) and marginalized over the other parameters.
    95\% credible upper limits on $\zeta_0$ in each bin are indicated as blue horizontal ticks, and the corresponding medians are indicated as black ticks.
    The dotted vertical line in the left panel corresponds to 1 yr.}
    \label{fig:upper-limit-freq-ecc}
\end{figure}

We converted the posterior samples from the detection analysis to posterior samples for the upper limit analysis (i.e., using a uniform prior on $\zeta_0$) using the sample reweighting method described in Ref. \cite{HourihaneMeyers+2022} and references therein. 
The two panels of Fig. \ref{fig:upper-limit-freq-ecc} show the posterior distribution of $\zeta_0$, binned in frequency and eccentricity respectively and marginalized over all other parameters, visualized using violin plots.
The 95\% credible upper limits on $\zeta_0$ in each frequency/eccentricity bin are also shown in Fig. \ref{fig:upper-limit-freq-ecc} using blue horizontal ticks.
Fig. \ref{fig:upper-limit-freq-ecc} shows that the dataset is not very sensitive to $f_{gw}$ below 10 nHz. 
The sensitivity is highest in the 10--17.78 nHz frequency bin and the corresponding 95\% credible upper limit on $\zeta_0$ is $\sim0.35$ $\mu$s.
Fig. \ref{fig:upper-limit-freq-ecc} also shows that the sensitivity does not vary as appreciably with eccentricity as with frequency for the dataset used.


\section{Summary and Discussion}
\label{sec:summary}

In this work, we have investigated the computational cost of four different methods of computing the PTA signal $R(t)$ given the orbital variables (the mean motion $n$, eccentricity $e$, mean anomaly $l$, and the periastron angle $\gamma$) as functions of time.
These four methods are (1) using approximate analytic expressions, (2) using a truncated Fourier series expansion, (3) using a post-circular expansion, and (4) using numerical integration.
Our performance comparison revealed that the approximate analytic expressions provide the best execution times out of the four methods, and are $\sim50$ times faster than numerical integration, which is the most accurate method.

We showed that the approximate analytic expressions incur an amplitude error at the 1PN level. 
We characterized this error by introducing a mismatch metric valid for comparing PTA signals.
This comparison revealed that the mismatch between the PTA signals computed using the analytic expressions and numerical integration is only $\sim0.01$ even for $M=10^9M_{\astrosun}$, $e_{t0}=0.85$ and $P_{b0}=0.5$ year;
i.e., the computationally efficient analytic expressions can be safely used within the region of the binary parameter space that is of interest to PTA experiments.

Although the analytic PTA signal expressions significantly improve the computational performance, they still require the evaluation of expensive hypergeometric functions for solving the orbital evolution.
To address this, we introduced a spline interpolation method that exploits the typical structure of the narrowband PTA datasets, where each epoch has multiple TOAs, to achieve further gains in computational performance.
This method computes $R(t)$ and $h(t)$ only once every epoch and interpolates the $R(t)$ values at every TOA using a cubic Hermite spline.
This method can in principle provide an improvement in performance by a factor equal to the average number of TOAs per epoch.

{To demonstrate the feasibility of a PTA search for eccentric SMBHB signals using the methods described in this work, we performed a single-pulsar Bayesian search for such signals using the NANOGrav 12.5-year narrowband data of PSR J1909-3744.
To this end, we derived a simplified PTA signal expression that removes the degeneracies that arise in the single-pulsar case.
Since the search did not yield any detections, we estimated 95\% credible upper limits on the PTA signal amplitude as functions of GW frequency and eccentricity.
While these upper limits are not stringent enough to be astrophysically relevant, they are nevertheless proof of concept for the feasibility of a PTA search using the methods presented in this work.}

The prohibitively high cost of computing the PTA signals has been a major bottleneck in searching for and constraining the presence of inspiralling eccentric SMBHBs in PTA datasets. 
The results presented in this paper will provide a significant boost to such endeavors.
These methods have been incorporated into the \GWeccjl{} package, and we plan to apply them to search for eccentric SMBHBs in the upcoming NANOGrav 15-year dataset and the IPTA Data Release 3. 

\section*{Data availability}

The NANOGrav 12.5-year dataset is available at \url{https://data.nanograv.org/}. 
{The MCMC chains from the single-pulsar Bayesian search are available at \url{https://doi.org/10.5281/zenodo.8028584}.}

\section*{Software}
\href{https://github.com/PumasAI/DataInterpolations.jl}{\texttt{DataInterpolations.jl}},
\href{https://github.com/JuliaDiff/FiniteDifferences.jl}{\texttt{FiniteDifferences.jl}},
\href{https://github.com/JuliaMath/HypergeometricFunctions.jl}{\texttt{HypergeometricFunctions.jl}},\\
\href{https://github.com/dextorious/NumericalIntegration.jl}{\texttt{NumericalIntegration.jl}},
\href{https://github.com/JuliaPhysics/PhysicalConstants.jl}{\texttt{PhysicalConstants.jl}},
\href{https://github.com/PainterQubits/Unitful.jl}{\texttt{Unitful.jl}},
\href{https://github.com/JuliaAstro/UnitfulAstro.jl}{\texttt{UnitfulAstro.jl}},
\texttt{numpy} \cite{HarrisMillman+2020}, 
\texttt{scipy} \cite{VirtanenGommers+2020},
\texttt{matplotlib} \cite{Hunter2007}, 
\texttt{GSL} \cite{GalassiDavies+2009}, 
\texttt{ENTERPRISE} \cite{EllisVallisneri+2020},
\texttt{enterprise\_extensions} \cite{TaylorBaker+2021}, 
\texttt{PTMCMCSampler} \cite{EllisvanHaasteren2017},
\texttt{corner} \cite{Foreman-Mackey2016}.

\begin{acknowledgments}
AS is supported by the NANOGrav NSF Physics Frontiers Center (awards \#1430284 and 2020265).
AS thanks Sarah Vigeland, David Kaplan, and Stephen Taylor for fruitful discussions, Amit Jit Singh, Nidhi Pant, Lankeswar Dey, and Belinda Cheeseboro for providing valuable feedback on the \GWecc{} package, and the anonymous referees for their detailed comments and suggestions on the manuscript.
\end{acknowledgments}

\appendix

\section{Number of harmonics required for accurate computation of quadrupolar-order gravitational waveforms using its Fourier series expansion}
\label{sec:pm-nharms}

\begin{figure*}
    \centering
    \includegraphics[scale=0.8]{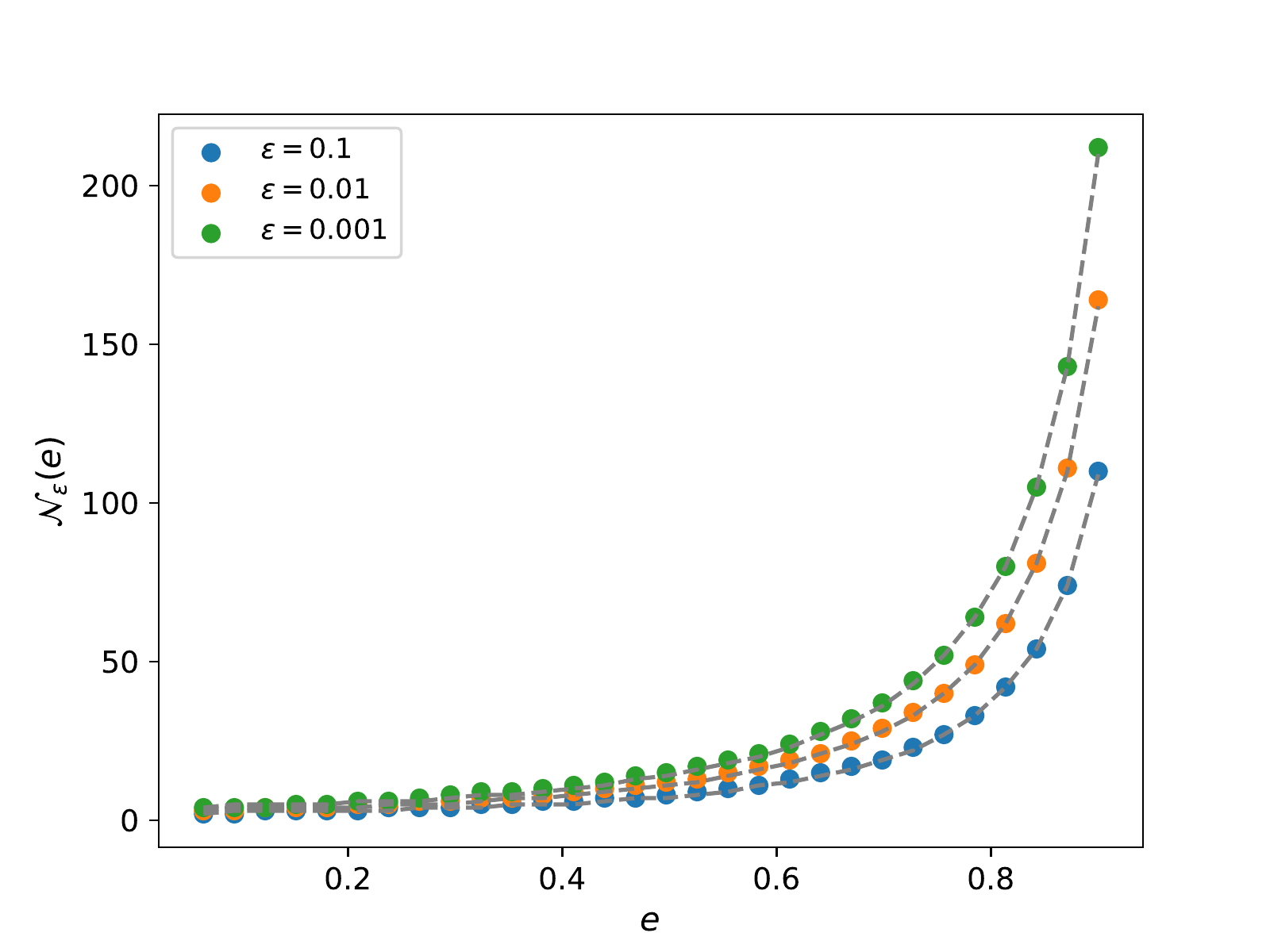}
    \caption{Number of harmonics $\mathcal{N}_\epsilon$ required to capture {the fraction $1-\epsilon$} of the total emitted power in a Peters-Mathews waveform.
    Circles represent numerically estimated $\mathcal{N}_\epsilon$ values {for different $\epsilon$} and the grey dashed curves represent the {corresponding fitting functions}.}
    \label{fig:nharms}
\end{figure*}

The total GW power emitted by an eccentric binary, accurate up to the leading quadrupolar order, is given by \cite{PetersMathews1963}
\begin{equation}
    P(e)=\frac{32}{5}\frac{c^{5}}{G}\left(\frac{GMn}{c^{3}}\right)^{10/3}\eta^{2}f(e)\,,
\end{equation}
where 
\begin{equation}
    f(e) = \frac{\left(1+\frac{73}{24}e^{2}+\frac{37}{96}e^{4}\right)}{\left(1-e^{2}\right)^{7/2}}\,,
\end{equation}
and the corresponding spectral power associated with the $p$th harmonic is given by \cite{PetersMathews1963}
\begin{equation}
    P_p(e) = \frac{32}{5} \frac{c^5}{G} \left(\frac{GMn}{c^3}\right)^{10/3} \eta^2 g_p(e)\,,
\end{equation}
where
\begin{widetext}
\begin{align}
g_{p}(e)&=\frac{p^{4}}{32}\left\{ \left[J_{p-2}(pe)-2eJ_{p-1}(pe)+\frac{2}{p}J_{p}(pe)+2eJ_{p+1}(pe)-J_{p+2}(pe)\right]^{2}\right.\nonumber\\
&\qquad\left.+\left(1-e^{2}\right)\left[J_{p-2}(pe)-2J_{p}(pe)+J_{p+2}(pe)\right]^{2}+\frac{4}{3p^{2}}\left[J_{p}(pe)\right]^{2}\right\}\,.
\end{align}
\end{widetext}
We define $\mathcal{N}_\epsilon(e)$ as the number of harmonics {that captures the fraction $1-\epsilon$} of the total emitted power for a given eccentricity $e$,
i.e., $\mathcal{N}_\epsilon(e)$ is the smallest integer such that
\begin{equation}
    \sum_{p=1}^{\mathcal{N}_\epsilon(e)} \frac{g_p(e)}{f(e)} \ge 1-\epsilon\,.
\end{equation}
We plot in Figure \ref{fig:nharms} numerically estimated values of $\mathcal{N}_\epsilon(e)$ for different values of $e$.

We have found that the following formula fits the numerically estimated $\mathcal{N}_\epsilon(e)$ well:
\begin{equation}
  \mathcal{N}_\epsilon(e) \approx \alpha_\epsilon (1-e^2)^{-3/2} + \beta_\epsilon\,, 
  \label{eq:nharms}
\end{equation}
{where $\alpha_\epsilon$ and $\beta_\epsilon$ are functions of $\epsilon$.
For instance, $\alpha_\epsilon=18.64801851$ and $\beta_\epsilon=-14.04695398$ for $\epsilon=10^{-3}$}.
The $\mathcal{N}_\epsilon(e)$ values predicted by this fitting formula are plotted in Figure \ref{fig:nharms} using a solid grey line.

\bibliographystyle{apsrev4-2}
\bibliography{pta-signal-adiabatic}

\end{document}